\documentclass[letterpaper, 10 pt]{IEEEtran}%
\usepackage{amsmath,amsfonts}

\usepackage[english]{babel}
\usepackage{array}
\usepackage[caption=false,font=normalsize,labelfont=sf,textfont=sf]{subfig}
\usepackage{textcomp}
\usepackage{stfloats}
\usepackage{url}
\usepackage{verbatim}
\usepackage{graphicx}
\usepackage{cite}
\makeatletter
\def\bibcite#1#2{\@ifundefined{b@#1}{\global\@namedef{b@#1}{#2}}{}}
\makeatother
\usepackage{units}
\usepackage{dsfont}
\usepackage{comment}
\usepackage{booktabs}
\usepackage{marginnote}
\usepackage{xfrac}
\usepackage{epstopdf}
\usepackage{algorithm}
\usepackage{algpseudocode}
\usepackage{circuitikz}
\usepackage{tikz}
\usepackage{tikz-uml}
\usepackage{multirow}
\usetikzlibrary{shapes, arrows,calc,positioning }
\usepackage{xcolor}%
\usepackage{bm}%
\usepackage{tabularx}
\usepackage{array}%

\definecolor{dataColor}{RGB}{0, 102, 204}
\definecolor{simColor}{RGB}{255, 94, 77}
\definecolor{infraColor}{RGB}{96, 143, 159}
\definecolor{opColor}{RGB}{34, 139, 34}

\usepackage[compact]{titlesec}
\titlespacing{\section}{10pt}{*1}{*1}
\titlespacing{\subsection}{8pt}{*0}{*0}
\titlespacing{\subsubsection}{4pt}{*0}{*0}
\usepackage{mathtools}
\usepackage{amsthm}

\newtheorem{theorem}{Theorem}[section]

\usepackage{color, colortbl}

\definecolor{expRR}{HTML}{7498C8}
\definecolor{expRO}{HTML}{FECB91}
\definecolor{expOR}{HTML}{A6D9A6}
\definecolor{expOO}{HTML}{F4766C}

\title{Battery Electric Truck Infrastructure Co-design via Joint Optimization and Agent-based Simulation}

\author{Juan Pablo Bertucci$^{1}$, Mauro Salazar$^{1}$ and Theo Hofman$^{1}$%
\thanks{This publication is part of the project GTD-Elektrifikatie, made possible in part by the Ministry of Economic Affairs and Climate Policy of the Netherlands. A preliminary version of this paper was presented at the 2024 American Control Conference.}%
\thanks{$^{1}$Department of Mechanical Engineering, Control System Technology,
        Eindhoven University of Technology, 5600 MB Eindhoven, The Netherlands
        {\tt\small \{j.p.bertucci, m.r.u.salazar, t.hofman\}@tue.nl}}%
}

\begin{document}

\maketitle
\thispagestyle{empty}
\pagestyle{empty}

\begin{abstract}
As zero-emission zones emerge in European cities, fleet operators are shifting to electric vehicles.
To maintain their current operations, a clear understanding of the charging infrastructure required and its relationship to existing power grid limitations is needed.
This study presents an optimization framework for jointly designing charging infrastructure and schedules within a logistics distribution network, validated through agent-based simulations.
We formulate the problem as a mixed-integer linear program and develop an agent-based model to evaluate various designs and operations under stochastic conditions.
Our experiments compare rule-based and optimized strategies in a case study of the Netherlands. Results show that current commercial solutions suffice for middle-mile logistics, with central co-design yielding average cost reductions of 5.2\% to 6.4\% and an average 20.1\% decrease in total installed power.
While rule-based control effectively manages charging operations and mitigates delays, optimizing charge scheduling significantly reduces queuing times (99\%), charging costs (13.5\%), and time spent near capacity (10.9\%).
Our optimization-simulation framework paves the way for combining optimized infrastructure planning and realistic fleet operations in digital-twin environments.
\end{abstract}
\section{Introduction}
Electrification of transportation is mandatory to reduce and the dependence on fossil fuels, and, in turn, improve air quality. Globally, land transportation accounts for nearly 40\% of CO$_2$ emissions \cite{teter2017future} and within the Netherlands for about 20\% of the total \cite{burgmeijer2018}. Within the transportation sector, road freight accounts for approximately 65\% of the world's freight emissions \cite{itf2021} and 20\% of the transportation greenhouse gas emissions in the Netherlands~\cite{cbsopendata}.
In response, the EU
has put forward an ambitious plan to foment the electrification of freight transport with the planned placement of Zero Emission Zones (ZEZs) in several major cities, where only zero-emission commercial vehicles will be allowed to operate.
Especially in the Netherlands, The Hague, Rotterdam, and Amsterdam and over 30 other cities intend to ban the use of commercial non-zero emissions vehicles in and near their downtown areas by 2025\cite{Rijksoverheid2021Nieuwe}.
Additionally, differential tax schemes and purchase investments have been introduced to encourage the purchase of electric trucks in the Netherlands, but also in Germany, Austria, Sweden, the United Kingdom and the United States \cite{Lantz.Joelsson2023}.
This has led to major incentives to transition to battery electric trucks (BETs) for freight, but the challenges of switching to electric trucks are manifold: the trade-off between range (battery size) and useful payload capacity; the higher purchasing costs compared to traditional diesel vehicles; the lack of charging infrastructure of adequate capacity in critical corridors; and uncertain provision of grid power (especially considering renewable sources) \cite{Ploetz2022,galloElectricTruckBus2016}.
All these aspects produce an adverse environment for a quick adoption of electric trucks in supply chains; not to mention possible operational complications due to coordinating the charging of vehicle fleets, whereas the lack of charging infrastructure has been recognized as one of the biggest bottlenecks in the widespread adoption of the technology \cite{Speth.etal2022a,Transport&Environment2023}.
Given the current circumstances, it is imperative to develop new models that guarantee efficient use of assets and energy to facilitate the transition to electric transportation. Optimization and simulation methods can help accelerate the transition by driving down design costs, while mitigating potential challenges and uncertainties faced by fleet operators that adopt these technologies.
In this paper, we devise a framework to jointly design the charging infrastructure and operations for a fleet of battery-electric trucks (BETs) from realistic delivery data, and a simulation environment to test its performance in stochastic conditions.
\begin{figure}[!t]
        \begin{centering}
            \includegraphics[width=0.9\linewidth]{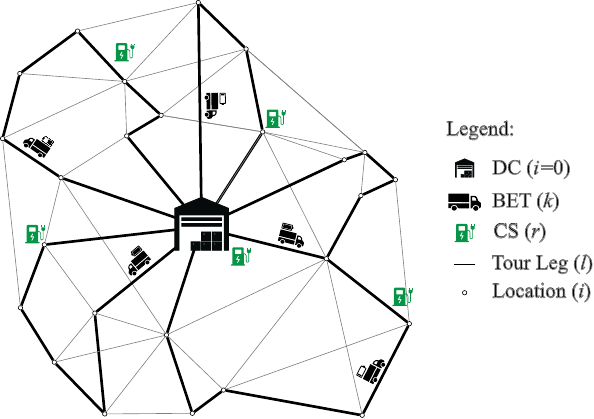}
        \caption{A network sketch where nodes represent retailers connected to a central distribution center (DC).
        This paper deals with the optimal location and sizing of charging stations (CS) for battery-electric truck (BET) operations.}
        \label{fig:intro-sketch}
        \end{centering}
        \end{figure}

\textit{Related Literature:} Our work contributes to two research streams, namely, placement and sizing of charging stations and scheduling of charging operations for BETs.
Infrastructure placement for BETs has been tackled with simulation, optimization and spatial anaylsis.
In \cite{Speth.etal2022a} BET adoption rates are assumed and using traffic count data paired with queuing models estimate the number of charging points and their locations for Germany.
In a similar fashion, \cite{Zhang.etal2021} uses spatial analysis combined with grid simulation models to estimate the optimal locations for BET charging.
In contrast, \cite{Whitehead.etal2022} combines spatial analysis of truck trip data with optimization, finding that a relatively modest public charging network should be sufficient to support the majority of short-haul electric trucks. They also conclude that the majority of short-haul electric truck charging is likely to be carried out at depots. This is also in agreement with the results from \cite{spethDepotSlowCharging2024}, who find slow depot charging is posed to have the greatest share of charging events in Germany using simulation.
Specifically, in \cite{NREL2023} an agent-based model (ABM) was developed to test the performance of individual truck charging stations.
This ABM model was combined with traffic spatial analysis in \cite{Mishra.etal2022} to obtain the requirements of a megawatt-level charging station, finding a mix of 1.2\unit{MW} and 100\unit{kW} chargers can achieve a good quality of service.
In the case of charging scheduling for BETs, both optimization and simulation methods have been employed.
Scheduling and routing is tackled by solving a mixed-integer linear program (MILP) for different combinations of BET and charger characteristics in \cite{pelletierChargeSchedulingElectric2018}.
Similarly, the authors of~\cite{zahringerTimeVsCapacity2022} employ gradient-free optimization and simulation to determine the optimal BET charging stop strategies considering a synthetic charging station network.
Also, \cite{zahringerOptimizingJourneyDynamic2024} uses dynamic programming and simulation to automate decision-making for re-charging stops in long-distance trips.
Later, \cite{bragin-joint} and \cite{wangOptimalDispatchRouting2023a} jointly optimize the routing, fleet size and charging schedule by solving a MILP assuming existing charging infrastructure locations. In~\cite{bragin-joint} the authors find that total costs do not decrease significantly for charging powers above 500\unit{kW}, for BETs with battery sizes of 250\unit{kWh}.
However, in all these cases the charging infrastructure is assumed or not co-designed with the fleet charging schedules.
Moreover, combining this with actual data for logistics and market-available chargers, and comparing operational and design costs with rule-based decision making has not been carried out in a simulation environment that considers queuing dynamics and operational stochasticity.

\textit{Statement of Contributions:} In this study, we combine realistic operational data with an optimization-simulation framework. We determine the optimal infrastructure and charging schedules, simulate the results in an agent-based simulation testing platform, and contrast the results obtained with optimization algorithms to myopic rule-based approaches for decision-making.
This contributes to the emerging literature on BET operations in multiple ways. First, we provide an applicable solution methodology for operators switching from diesel to electric trucks: They can design their charging infrastructure based on their current logistic needs and with optimality guarantees.
Second, we test our results against a rule-decision approach, illustrating the value of using jointly optimized charging schedules and infrastructure for fleet operators, not only in terms of costs but also in terms of grid requirements.
Third, we combine optimization and simulation to test the robustness of the solutions obtained with optimization; using a training and testing set from a real-world scenario we analyze how each solution would perform under conditions outside the values used for optimization, with stochastic perturbations.
A preliminary version of this paper was presented in~\cite{BertucciHofmanEtAl2024}.
In this extended version, we equip the approach with a simulation model, modify the optimization problem to include additional practical charging constraints and modify the decision variables to allow for continuous power values in charging operations, and conduct a stochastic analysis of the solutions.
Furthermore, we quantify the value of using co-designed infrastructure and using optimized charging scheduling by comparing results of rule-based to optimized decision making.

\textit{Organization:} The remainder of this paper is structured as follows:
In Section \ref{sec:meth}, we present the methods used to distinguish between the initial optimization and the subsequent simulation procedure.
In Section \ref{sec:case} we outline the main variables and experiments deployed in the case study for a distribution center in North Holland.
In Section \ref{sec:res}, we present the results of the case study,
and in Section \ref{sec:dis} we analyse the main conclusions for our case study and give future research directions.

\definecolor{dataColor}{HTML}{F5F5F5}
\definecolor{infraColor}{HTML}{F5F5F5}
\definecolor{opColor}{HTML}{F5F5F5}
\definecolor{simColor}{HTML}{F5F5F5}
\definecolor{arrowColor}{HTML}{000000}

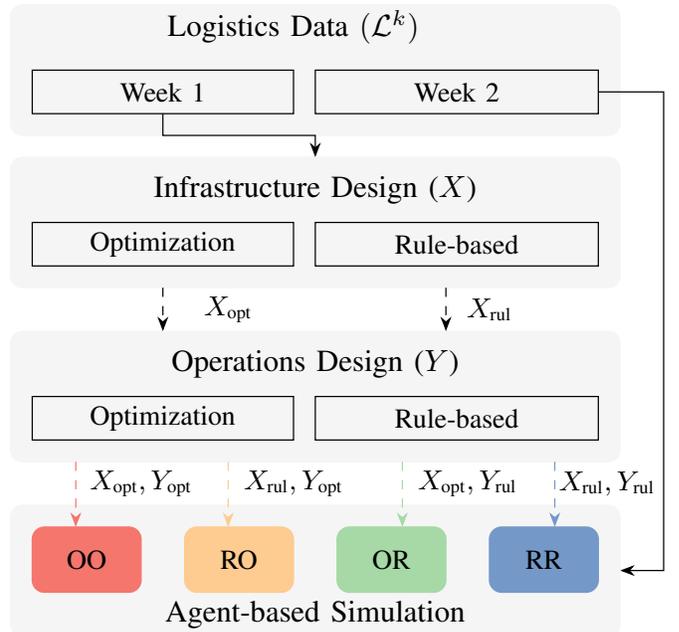
\begin{figure}[!t]
    \centering
    \resizebox{1\linewidth}{!}{%
        \begin{circuitikz}
        \tikzstyle{every node}=[font=\small]

        \tikzstyle{dataRect} = [draw=dataColor, fill=dataColor,  rounded corners]
        \tikzstyle{simRect} = [draw=simColor, fill=simColor,  rounded corners]
        \tikzstyle{infraRect} = [draw=infraColor, fill=infraColor,  rounded corners]
        \tikzstyle{opRect} = [draw=opColor, fill=opColor,  rounded corners]
        \tikzstyle{ooRect} = [draw=expOO, fill=expOO, rounded corners]
        \tikzstyle{orRect} = [draw=expOR, fill=expOR,  rounded corners]
        \tikzstyle{roRect} = [draw=expRO, fill=expRO,  rounded corners]
        \tikzstyle{rrRect} = [draw=expRR, fill=expRR, rounded corners]
        \tikzstyle{weekRect} = [draw=black]

        \draw[dataRect] (3,12.25) rectangle (10,10.75);

        \draw[infraRect] (3,10.5) rectangle (10,9);

        \draw[opRect] (3,8.5) rectangle (10,7);
        \draw[weekRect] (3.25,7.75) rectangle node { Optimization} (6.25,7.25);
        \draw[weekRect] (6.5,7.75) rectangle node { Rule-based} (9.75,7.25);

        \draw[weekRect] (3.25,9.75) rectangle node { Optimization} (6.25,9.25);
        \draw[weekRect] (6.5,9.75) rectangle node {Rule-based} (9.75,9.25);
        \draw[weekRect] (9.75,11.5) rectangle node { Week 2} (6.5,11);
        \draw[weekRect] (3.25,11.5) rectangle node { Week 1} (6.25,11);

        \draw [ color={arrowColor}, ->, >=Stealth] (9.75,11.25) -- (10.5,11.25) -- (10.5,5.75) -- (10,5.75);
        \draw [ color={arrowColor}, ->, >=Stealth] (4.75,11) -- (4.75,10.75) -- (6.5,10.75)-- (6.5,10.5);

\draw [ color={arrowColor}, ->, >=Stealth, dashed] (8,9) -- (8,8.5);
\draw [ color={arrowColor}, ->, >=Stealth, dashed] (4.75,9) -- (4.75,8.5);
\node [font=\small] at (5.5,8.75) {$X_{\text{opt}}$};
\node [font=\small] at (8.5,8.75) {$X_{\text{rul}}$};

\draw[simRect] (3,6.5) rectangle (10,5);
\draw[ooRect] (3.25,6.25) rectangle node { OO} (4.5,5.5);
\draw[orRect] (6.75,6.25) rectangle node { OR} (8,5.5);
\draw[roRect] (5,6.25) rectangle node { RO} (6.25,5.5);
\draw[rrRect] (8.5,6.25) rectangle node { RR} (9.75,5.5);

\draw [ color=expRR, ->, >=Stealth, dashed] (9.25,7) -- (9.25,6.25);
\draw [ color=expOO, ->, >=Stealth, dashed] (3.75,7) --  (3.75,6.25);
\draw [ color=expOR, ->, >=Stealth, dashed] (7.5,7) --  (7.5,6.25);
\draw [ color=expRO, ->, >=Stealth, dashed] (5.5,7) --  (5.5,6.25);

\node [font=\small] at (4.5,6.75) {$X_{\text{opt}},Y_{\text{opt}}$};

\node [font=\small] at (6.25,6.75) {$X_{\text{rul}},Y_{\text{opt}}$};

\node [font=\small] at (8.25,6.75) {$X_{\text{opt}},Y_{\text{rul}}$};

\node [font=\small] at (9.85,6.75) {$X_{\text{rul}},Y_{\text{rul}}$};

        \node [font=\normalsize] at (6.25,12) {Logistics Data $(\mathcal{L}^k)$};

        \node [font=\normalsize] at (6.5,10.125) {Infrastructure Design ($X$)};

        \node [font=\normalsize] at (6.5,8.125) {Operations Design ($Y$)};

        \node [font=\normalsize] at (6.5,5.25) {Agent-based Simulation};

        \end{circuitikz}
    }%

    \caption{Flowchart depicting the overarching methodology proposed. Input data on existing operations is given for each vehicle trip as the set $\mathcal{L}^k$. Then, through the optimization module, the optimal design variables $X_i^r$ (quantity, location and power rating of chargers) and $Y^{r,t}_{k,l}$ (charging schedule of each truck) are obtained. Alternatively, the optimization module may be bypassed and rule-based decision making is applied to obtain the charging infrastructure and charging schedules are defined in-simulation.
   }
    \label{fig:methodology-process-flow}
\end{figure}

\section{Methodology}\label{sec:meth}

This section describes our proposed methodology outlined in Fig.~\ref{fig:methodology-process-flow}.
First, the initial data and mathematical sets required for the optimization formulation (based on real-world itinerary data of a logistic operator) are described.
Second, we detail the MILP formulation that determines the optimal number of chargers at each distribution center and the charging schedule of each vehicle.
Third, we develop our simulation environment, where the different components of the system (vehicles, retailers, and distribution centers) are modeled as individual objects in an agent-based model.
Finally, we define the operational metrics that will be applied to asssess the robustness of the designs using Monte Carlo simulations.
\subsection{Data Inputs}
The main data inputs for the problem comprise, first,
the vehicle fleet data which characterize the $k \in \mathcal{K}$ vehicles that will make use of the charging infrastructure, with their respective battery size $E_k$ (\unit{kWh}), truck weight $w_k$ (\unit{ton}), consumption per \unit{ton-km} driven $e^{\mathrm{ton.km}}_k$ (\unitfrac{kWh}{ton-km}), auxiliary consumption per hour $c^{\mathrm{aux}}_k$ ($\unitfrac{kWh}{hr}$) and cooling consumption $c^{\mathrm{cool}}_k$ $(\unitfrac{kWh}{km})$.
Second, the location data set comprises the origins and destinations represented as $i,j \in \mathcal{I}$, where the distance between each pair $s_{ij}$ is given. For each location we will also have a connection cost related to the peak power used, $C^\mathrm{peak}_i$ (€$\unitfrac{}{kW}$), and the energy price for each location $i$ and for each time step $t \in \mathcal{T}$, $p^t_i$ (€$\unitfrac{}{kW}$).
Third, the chargers to be used in the design will be denoted by their type $r \in \mathcal{R}$, and will have corresponding maximum power $e^{\mathrm{char}}_r$ ($\unit{kW}$), efficiencies $\eta_r$ (dimensionless) and costs $C^r$ (€).
Finally, the itineraries to be undertaken by each truck $k$ are represented with the set $\mathcal{L}^k$, where we use $l_k \in \mathcal{L}_k$ to denote each specific trip leg of each truck $k$.
For each trip leg we will have an origin $i_l$, a destination $j_l$, a distance $s_l$ (determined by the pair $(i_l,j_l)$), a payload to be carried $w_l$ ($\unit{ton}$), an indicator variable for refrigeration requirements $b^\mathrm{cold}_l$, as well as time windows for the departure $t^{\mathrm{dep.ear}}_l$, $t^{\mathrm{dep.lat}}_l$ and arrival times $t^{\mathrm{arr.ear}}_l$, $t^{\mathrm{arr.lat}}_l$. These time windows delimit the earliest and latest time for each vehicle to depart from its current location and arrive to the destination location, and are given in the daily delivery planning of the operator.
Travel times $t^\mathrm{travel}_{l}$, loading times $t^\mathrm{load}_{l}$ and unloading times $t^\mathrm{unload}_{l}$ ($\unit{hours}$) are also part of the itinerary information provided (or calculated on the basis of other considerations).
\label{sec:meth:intro}

\subsection{Mixed Integer Optimization Problem}\label{section:opt}
\begin{figure*}[!t]
\begin{centering}
\includegraphics[width=\linewidth,trim={5mm 48mm 5mm 48mm},clip]{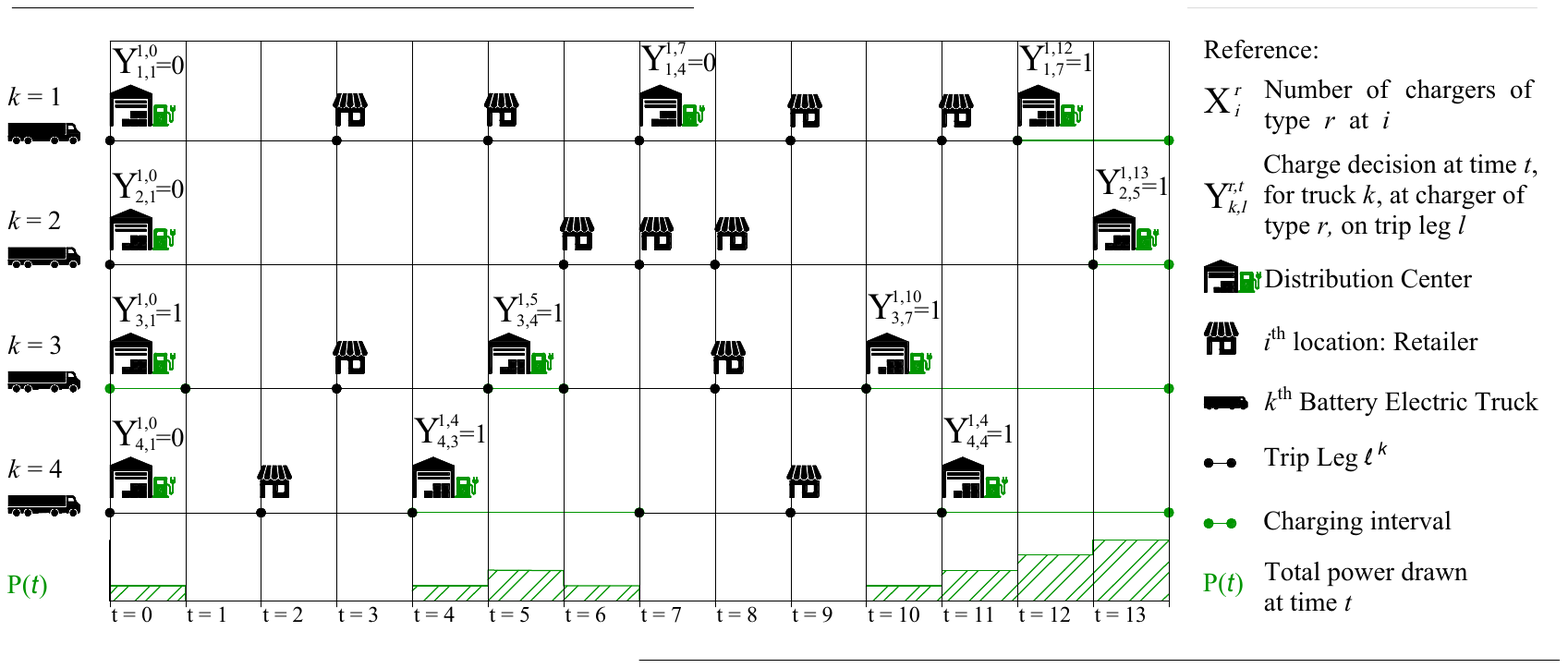}
\caption{
Schematic of the optimization problem solved and its main decision variables:
$k$ BETs have pre-specified delivery tours comprised of multiple legs $l_k$ that span between locations (DCs or retailers) denoted by $i$ (origin) and $j$ (destination).
The number of chargers of type $r$ available at each location $i$ is given by the integer decision variable $X^{r}_{i}$.
BETs have the decision to charge before each trip leg $l_k$, at a given time step $t$, at a charger of type $r$. This is modelled with a binary decision variable defined as $Y^{k,l}_{r,t}$.
These decisions add up to a final power requirements curve denoted as $P(t)$.
For simplicity's sake, in this diagram chargers are only installed at the DC (i.e: $X^r_{i\neq1}=0$).
}
\label{fig:optimization-schematic}
\end{centering}
\end{figure*}

We now develop the optimization formulation to jointly determine the required infrastructure and optimal charging schedules of a BET fleet for a given set of deliveries to be completed.
We discretize the total simulation time (spanning multiple days) into $\mathcal{T}$ units of a size $\tau$ (e.g., fraction of hour), such that the whole period of analysis can be described through $t \in \mathcal{T}$.
We use $l$ to index each element (trip leg) in an itinerary $\mathcal{L}_k$; thus $l_k$ represents the $l$-th element in the set $\mathcal{L}_k$ (where $l = 1, 2, ..., |\mathcal{L}_k|$). Therefore, each index combination $(l,k)$ identifies a specific trip by a specific vehicle, with all its characteristics given in \ref{sec:meth:intro}.
\subsubsection{Decision Variables}
We allow the vehicles to charge before each departure for a given trip leg $l$. We then define the choice of a truck $k$ to charge before starting itinerary leg $l$ at a charger of type $r$ at a time slot $t$ as a binary variable $Y^{r,t}_{k,l}$. This decision can be taken prior to each departure, to fulfill the energy requirements the upcoming trip, and respecting the time windows for each leg.
The actual times of departure and arrival are given by the decision variables $t^{\mathrm{dep.act}}_{k,l}$ and $t^{\mathrm{arr.act}}_{k,l}$, which have to be consistent with the chosen charging intervals.
In a similar fashion, we define the continuous variable $P^{t}_{k,l}$ which is the actual power delivered to a truck $k$ before completing itinerary leg $l$ in a time slot $t$ (i.e., corresponding to each decision variable $Y^{r,t}_{k,l}$). This value will be limited by the maximum rated power of the charger $e^{\mathrm{char}}_r$.
Whether trucks can effectively charge at a location depends on the existing infrastructure, namely, the number of chargers of the required type $r$ at the departure location $i$, denoted as $X^r_i$.
\subsubsection{Constraints}
The vehicles are subject to energy requirements, infrastructure availability, and schedule constraints:
\textit{Energy Requirements:} We require the vehicles to have enough energy at each origin to arrive at their next destination. To guarantee this, we specify the state of energy (SoE) at each destination $j$ of each trip leg $l$ to be at least the energy at the origin $i$ minus the energy needed to travel distance $s^k_{l}$, with payload $w^k_l$.
We then define the energy consumed in carrying out each leg by
\begin{equation}
\label{eq:energy-req-veh}
    E^{\mathrm{cons}}_{k,l}  = e^{\mathrm{ton.km}}_k \, (s^k_l \, w^\mathrm{tot}_{k,l}) + (c^{\mathrm{aux}}_k + c^{\mathrm{cool}}_k\, b^\mathrm{cold}_l)\, t^\mathrm{travel}_{l},
\end{equation}
where $w^\mathrm{tot}_{k,l}$ is the sum of the chassis and battery weight of the truck $w_k$ and the payload specific to the current leg $w_{k,l}$.
To complete a trip $l_k$, a truck may charge at a power rating $P^{t}_{k,l}$ if there is a charger available at the leg origin. We define the energy charged at the beginning of a leg $l_k$ as
\begin{equation}
\label{eq:energy-charged-pre-trip}
    E^{\mathrm{char}}_{k,l}  = \tau \, \sum_{t \in T^\mathrm{sub}_{k,l}} P^{t}_{k,l} ,
\end{equation}
where the subset $T^\mathrm{sub}_{k,l}$ is defined by the time blocks $t$ that lie between the previous earliest possible arrival time $t^{\mathrm{arr.ear}}_{k,l-1}$ and latest possible departure time $t^{\mathrm{dep.lat}}_{k,l} + \beta^{\mathrm{slack}}$.
We add the tuning parameter $\beta^{\mathrm{slack}}$ to give some additional time slack to the original schedules, given that the original itineraries are derived from the vehicle routing of internal combustion engine vehicles, and these could be too restrictive for the charging times required for a specific BET trip.
The charging power chosen $P^{t}_{k,l}$ is limited by the maximum charger power in the following constraint:
\begin{equation}
    \label{eq:ctr:energy-charged-leq-choice-max}
    P^{t}_{k,l} \leq \sum_{r \in R} Y^{r,t}_{k,l}\, e^{\mathrm{char}}_r
    \quad \forall l \in \mathcal{L}_k,
    \forall k \in \mathcal{K},
    \forall t \in \mathcal{T}.
\end{equation}
With this set-up, we can then write the first inequalities that guarantee the energy feasibility of a trip leg $l_k$ as
\begin{equation}
\label{eq:ctr:energy-leg-update}
E^\mathrm{min}_k  \leq E^{\mathrm{dep}}_{k,l}
    + E^{\mathrm{char}}_{k,l}
    - E^{\mathrm{cons}}_{k,l} \leq E^{\mathrm{arr}}_{k,l}
\quad \forall l \in \mathcal{L}_k,
\forall k \in \mathcal{K},
\end{equation}
where we limit the energy not to drop below a minimum level $E^\mathrm{min}_k$ that acts as a buffer in case of variations in $E^{\mathrm{cons}}_{k,l}$.
To have a matching state of charge at the end of a leg, and at the beginning of the subsequent one, we have the following equality:
\begin{equation}
    \label{eq:ctr:energy-legs-consistency}
        E^{\mathrm{arr}}_{k,l-1} = E^{\mathrm{dep}}_{k,l}
    \quad \forall l \in \mathcal{L}_k,
    \forall k \in \mathcal{K}.
    \end{equation}
Additionally, we limit the energy charged to be supported by the vehicle battery $E^\mathrm{max}_k$ at each trip leg,
\begin{equation}
\label{eq:ctr:battery-cap}
E^{\mathrm{dep}}_{k,l} + E^{\mathrm{char}}_{k,l} \leq E^\mathrm{max}_k
\quad \forall l \in \mathcal{L}_k,
\forall k \in \mathcal{K}.
\end{equation}

\textit{Schedules:} To obtain charging schedules that are coherent with the final logistic schedules, we need to enforce that the time of departure is posterior to the latest charging block of trip leg $l$. This is enforced as
\begin{equation}
    \label{eq:ctr:time-coherence-dep-char}
    t \, \sum_{r\in R} Y^{r,t}_{k,l} \leq   t^{\mathrm{dep.act}}_{k,l}
    \quad \forall t \in T^\mathrm{sub}_{k,l},
    \forall k \in \mathcal{K},
    \forall l \in \mathcal{L}_k.
\end{equation}
Correspondingly, the chosen departure time and travel time to the next destination must be consistent with the time window allowed between the actual arrival time from the previous leg $(l-1)$ and the latest departure time during the current leg $l$,
\begin{equation}
    \begin{split}
    \label{eq:ctr:time-coherence-dep-arr-next}
    t^{\mathrm{dep.act}}_{k,l-1} + t^\mathrm{travel}_{k,l-1} \leq t^{\mathrm{arr.act}}_{k,l} -\beta^{\mathrm{slack}}
\\ \quad%
    \forall k \in \mathcal{K},
    \forall l \in \mathcal{L}_k.
    \end{split}
\end{equation}
Analogously, considering the loading and unloading times:
\begin{equation}
    \begin{split}
    \label{eq:ctr:time-coherence-dep-arr-next-loading}
     t^{\mathrm{arr.act}}_{k,l-1} + t^\mathrm{load}_{k,l-1} + t^\mathrm{unload}_{k,l-1} \leq t^{\mathrm{dep.act}}_{k,l}
      \leq t^{\mathrm{dep.lat}}_{k,l} + \beta^{\mathrm{slack}}
\\ \quad%
    \forall k \in \mathcal{K},
    \forall l \in \mathcal{L}_k.
    \end{split}
\end{equation}

\textit{Infrastructure Availability:} The number of vehicles charging simultaneously is limited by the total number of available chargers at the departure location. Thus we write the constraint where for each location $i$ and for each type of charger $r$, we force the total number of vehicles charging at a time interval $t$ to be less than the number of available chargers $X_i^r$:
\begin{equation}
    \label{eq:ctr:char-ctr}
    \sum_{k\in K, l \in \mathcal{L}_k} Y^{r,t}_{k,l} \leq X_i^r
    \quad \forall t \in \mathcal{T},
    \forall r \in \mathcal{R},
    \forall i \in \mathcal{I}.
\end{equation}
We also ensure that a vehicle does not charge simultaneously at multiple chargers through
\begin{equation}
    \label{eq:ctr:vehs-unique-char}
    \sum_{r \in \mathcal{R}} Y^{r,t}_{k,l}  \leq 1
    \quad \forall t \in \mathcal{T},
    \forall k \in \mathcal{K},
    \forall l \in \mathcal{L}_k.
\end{equation}

\textit{Practical Charging Constraints:} To ensure that charging schedules do not involve switching between chargers in consecutive time blocks, we add the following auxiliary constraints:
\begin{equation}
    \begin{split}
    \label{eq:aux_vars_uv}
    v^{t}_{k,l} \leq Y^{r,t-1}_{k,l} - Y^{r,t}_{k,l} \leq u^{t}_{k,l},
\\    \quad \forall t \in \mathcal{T},
    \forall k \in \mathcal{K},
    \forall r \in \mathcal{R},
    \forall l \in \mathcal{L}_k,
    \end{split}
\end{equation}
where $u^{t}_{k,l}$ and $v^{t}_{k,l}$ are auxiliary binary variables. By adding the constraint
\begin{equation}
    \label{eq:aux_vars_uv_difference}
    u^{t}_{k,l} - v^{t}_{k,l} \leq 1,
    \quad \forall t \in \mathcal{T},
    \forall k \in \mathcal{K},
    \forall l \in \mathcal{L}_k,
\end{equation}
we enforce the charging choices transitions between two succesive time blocks to be either between no-charging and charging, or to continue charging in the same type $r$.

\textit{Energy Periodicity Constraints:} We assume a state of charge value for the trucks' batteries of $\phi$ at the beginning and end of the analysis period, to account for previous and posterior operations to our analysis timespan, where we denote our last leg as $l^\mathrm{max}_k$
\begin{equation}
\label{eq:ctr:periodicity-start}
E^{\mathrm{arr}}_{k,l=1} \leq \phi \, E^\mathrm{max}_k
\quad \forall k \in \mathcal{K}.
\end{equation}
\begin{equation}
\label{eq:ctr:periodicity-end}
\phi \, E^\mathrm{max}_k \leq E^{\mathrm{arr}}_{k,l=l^\mathrm{max}_k}
\quad \forall k \in \mathcal{K}.
\end{equation}

\subsubsection{Objective}

We define our cost function as the combination of energy, infrastructure, and peak consumption costs.

\textit{Energy Costs:} The time spent charging for each vehicle is given by the time step $\tau$ constant multiplied times the number of time blocks $t$ used for charging. The charging costs are obtained in a similar fashion as the energy charged given at~\eqref{eq:energy-charged-pre-trip}; they are obtained as the amount of power delivered by the charger (given by the power of charge in~\eqref{eq:energy-charged-pre-trip}) multiplied by the energy costs $p^t_i$ for each time interval and location. We obtain this value for each truck in the fleet,
\begin{equation}
    \label{eq:cost:charging}
     C^{\mathrm{char}}_{k, i} =  \tau \, \sum_{t\in T_\mathrm{sub}^{l,k}}%
     \frac{P^{t}_{k,l}}{\eta^r} \, p^t_i.
\end{equation}

\textit{Infrastructure Costs:} The infrastructure costs are given by the number of chargers built at each location $X^{r}_{i}$ multiplied by the capital cost of each charger of type $C^r$ :
\begin{equation}
    \label{eq:cost:infra}
     C^{\mathrm{infra}}_{i} =  \sum_{r \in \mathcal{R}} C^r \, X^{r}_{i}.
\end{equation}

\textit{Peak Consumption Costs:} We include the connection costs of each location, these are proportional to the peak (maximum) power usage during our period of analysis $t \in \mathcal{T}$ for every location $i \in \mathcal{I}$:
\begin{equation}
    \label{eq:cost:peak}
     C^{\mathrm{peak}}_{i} = \max_{t\in\mathcal{T}} \left\{ \tau \,\sum_{k \in K, l \in \mathcal{L}_k, r \in \mathcal{R}} \, Y^{r,t}_{k,l} \, e^{\mathrm{char}}_r \, c^\mathrm{peak}\right\},
\end{equation}
where $c^\mathrm{peak}$ corresponds to the prorrated cost for the peak power used during the entire period of analysis.
To keep the formulation as a linear problem, we relax the equality~\eqref{eq:cost:peak} with the following inequality:
\begin{equation}
    \label{eq:ctr:cost:peak}
     C^{\mathrm{peak}}_{i} \geq \sum_{k \in K,  l \in \mathcal{L}_k, r \in \mathcal{R}} Y^{r,t}_{k,l} \, e^{\mathrm{char}}_r \, c^\mathrm{peak} \quad \forall i \in \mathcal{I}, \forall t \in \mathcal{T}.
\end{equation}
Since we will be minimizing the variable $C^{\mathrm{peak}}_{i}$ in the objective, the inequality in~\eqref{eq:cost:peak} will hold with equality at the solution~\cite{BorsboomFahdzyanaEtAl2021}.

Finally, we define the logistic operator's optimization problem to be centered around minimizing the total costs of the operation. This is comprised of the global operation costs (charging and peak costs) and facility costs (location, number and type of chargers). Combining~\eqref{eq:cost:charging},~\eqref{eq:cost:infra} and~\eqref{eq:cost:peak} (multiplied by a parameter $\alpha^\mathrm{peak}$, a tuning parameter representing a possible increase in peak power costs) we define our objective function as
\begin{equation}
    \begin{aligned}
        \label{eq:obj-def}
J(X_i^r, Y^{r,t}_{k,l}, P^{t}_{k,l}) = \gamma^\mathrm{energy} \sum_{k \in K, i \in I}  C^{\mathrm{char}}_{k, i}
 + \sum_{i\in I}C^{\mathrm{infra}}_{i} \\
+ \alpha^\mathrm{peak} \, \sum_{i \in I}C^{\mathrm{peak}}_{i}.
    \end{aligned}
\end{equation}

\subsubsection{Problem Formulation}
We can thus write the minimization of the operator's cost as follows:

\textbf{Problem 1} (Joint Infrastructure Design and Charge Scheduling)
\textit{Given a set of vehicles $k$, with itineraries $L_k$, the number and power rating of chargers $(X_i^r$) and charging schedules ($Y^{r,t}_{k,l}, P^{t}_{k,l}$) minimizing total cost ($J$) result from}

\begin{equation}
\begin{aligned}
\label{eq:final-prob}
\min_{X_i^r, Y^{r,t}_{k,l}, P^{t}_{k,l}} \quad & J(X_i^r, Y^{r,t}_{k,l}, P^{t}_{k,l}) \\
\textrm{s.t.} \quad &
~\eqref{eq:energy-charged-pre-trip}-\eqref{eq:cost:infra}, ~\eqref{eq:ctr:cost:peak}, ~\eqref{eq:obj-def} \\
& Y^{r,t}_{k,l} \in \{0,1\}  \quad \forall r \in \mathcal{R},  t \in \mathcal{T}, k \in \mathcal{K}, l \in \mathcal{L}_k  \\
& X_i^r \in \mathbb{N} \quad \forall r \in \mathcal{R}, i \in \mathcal{I}.
\end{aligned}
\end{equation}
Problem 1 is a MILP that can be solved with global optimality guarantees with off-the-shelf optimization algorithms \footnotemark. The results from this problem are then added to the existing data structures from \ref{sec:meth:intro} and used in the simulation environment developed below, and contrasted in the experiments outlined in Fig.~\ref{fig:methodology-process-flow}.
\footnotetext{The code corresponding to this paper can be found in https://github.com/JPchomp/depotCharInf-public}
\tikzumlset{fill class=gray!10, fill package=white}
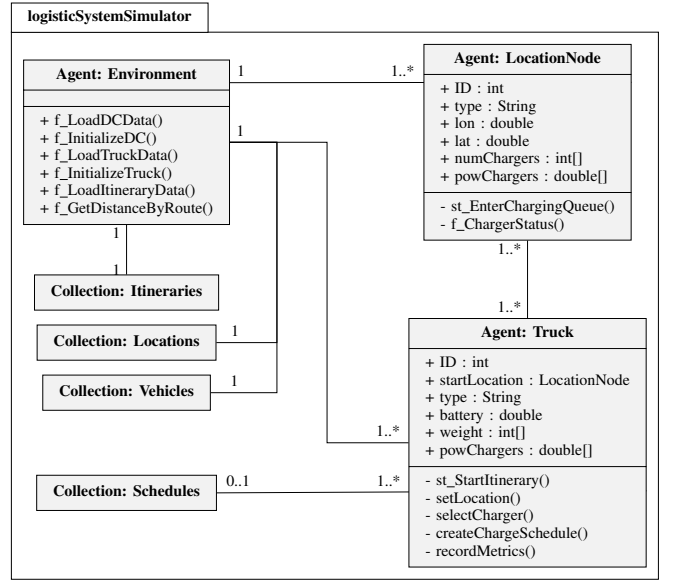
\begin{figure}[t!]
    \centering
    \resizebox{\columnwidth}{!}{
\begin{tikzpicture}

\begin{umlpackage}{logisticSystemSimulator}

\umlsimpleclass[x = 0, y = -3]{Collection: Itineraries}
\umlsimpleclass[x = 0, y = -4]{Collection: Locations}
\umlsimpleclass[x = 0, y = -5]{Collection: Vehicles}
\umlsimpleclass[x = 0, y = -7]{Collection: Schedules}

\umlclass[x = 0, y = 0]{Agent: Environment}{
}{
+ f\_LoadDCData() \\ + f\_InitializeDC() \\
+ f\_LoadTruckData() \\ + f\_InitializeTruck() \\
+ f\_LoadItineraryData() \\ + f\_GetDistanceByRoute()
}
\umlclass[x = 8, y=0]{Agent: LocationNode}{
  + ID : int \\ + type : String \\
  + lon : double \\ + lat : double \\
  + numChargers : int[] \\ + powChargers : double[]
}{
  - st\_EnterChargingQueue() \\
  - f\_ChargerStatus()}
\umlclass[x = 8, y=-6]{Agent: Truck}{
  + ID : int \\ + startLocation : LocationNode \\
  + type : String \\ + battery : double \\
  + weight : int[] \\ + powChargers : double[]
}{
  - st\_StartItinerary() \\ - setLocation() \\
  - selectCharger() \\ - createChargeSchedule() \\ - recordMetrics()
}

\umlassoc[geometry=--, mult1=1..*, pos1=0.57,  mult2=1, pos2=0.972,anchors=30 and 30]{Agent: LocationNode}{Agent: Environment}

\umlassoc[geometry=-|-,mult1=1..*, pos1=0.25, mult2 = 1, pos2 = 2.9]{Agent: Truck}{Agent: Environment}

\umlassoc[geometry=--, mult1=1..*, pos1=0.1 ,mult2 = 1..*, pos2 = 0.85 ]{Agent: LocationNode}{Agent: Truck}

\umlassoc[geometry=-- ,mult1=1..*,pos1=0.6, mult2 = 0..1, pos2 = 0.95,anchors=-22 and 0]{Agent: Truck}{Collection: Schedules}

\umlassoc[geometry=--, mult1=1, pos1=0.15 ,mult2 = 1, pos2 = 0.9]{Agent: Environment}{Collection: Itineraries}

\umlassoc[geometry=-|-,arm1 = 3, mult2 = 1, pos2 =2.7]{Agent: Environment}{Collection: Locations}

\umlassoc[geometry=-|-,arm1 = 3, mult2 = 1, pos2 =2.65 ]{Agent: Environment}{Collection: Vehicles}

\end{umlpackage}
\end{tikzpicture}
}
\caption{Unified modeling language class diagram for the agent-based simulation model. Four main collections are given to the model: the truck itineraries, the truck charging schedules, the fleet characteristics and the distribution center characteristics. The environment then initializes the DC agents and truck agents in their geographical location and the metrics resulting from the interactions are recorded.}
\label{fig:uml}
\end{figure}

\subsection{Agent-based Model}\label{section:abm}
We develop an agent-based simulation on the Anylogic platform \cite{borshchev2014multi}, where the different components of the system are modeled as individual objects interacting in a shared environment, as depicted in Fig.~\ref{fig:uml}. In essence, we simulate the main actors of this logistic system: vehicles and facilities. Both exist in a common spatial environment and operate under a shared simulation clock. We outline the main characteristics and behaviors of each agent type, the parameters and distributions used to add stochasticity in the simulations, and the final metrics recorded.

\subsubsection{Vehicles}
\begin{figure}[!t]
\begin{centering}
\includegraphics[width=\columnwidth]{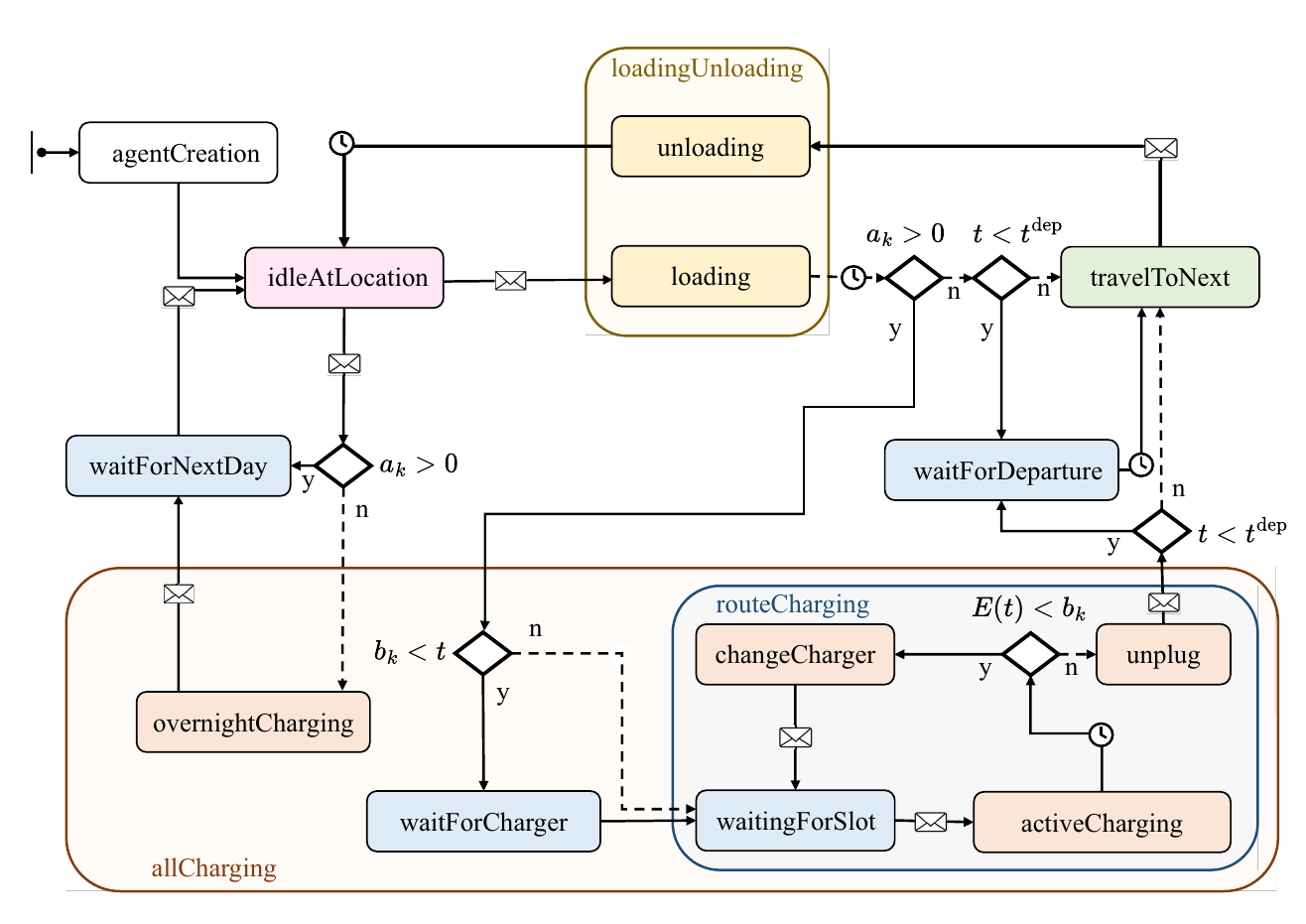}
\caption{Activity diagram for a truck $k$. This statechart can be classified in three operational states: Charging, Loading/unloading and Transport. In the case of the optimized schedule testing, the charging time and power is given in a separate charging schedule. In the rule-based case, the charging is performed depending on the instantaneous battery level.}
\label{fig:st_truck}
\end{centering}
\end{figure}

The vehicle agents represent the BETs that will carry out the transportation of goods from distribution center to retailers. These receive a collection of daily itineraries planned (as seen in Fig.~\ref{fig:uml}), consisting of the origins, destinations and payloads to be transported along with the arrival and departure schedules. We model the different state and decision-making of the trucks according to the activity diagram shown in Fig.~\ref{fig:st_truck}. Each truck $k$ successively performs loading, charging, transport, unloading, and overnight charging operations according to their respective itineraries, all according to the data outlined in \ref{sec:meth:intro}.
The vehicle agents are endowed with the decision-making capability of charging before starting a new itinerary according to their SoE at time step $t$. We denote the charging policy of a truck $k$ with $\pi_k=[a_k,b_k,c_k,d_k]$, where $a_k$ is the total energy to be charged, $b_k$ is the time to start charging, $c_k$ is the charger type to be used and $d_k$ is the power to be drawn. The policy $\pi_k$ results from the concatenation of four specific policies.
For the case of the amount of energy to be charged $a_k$
\begin{equation}
    \label{eq:abm2}
    a_k =
    \begin{cases}
    0, & \text{if } E_k(t) \geq E^\mathrm{cons}_{k,l} + E^\mathrm{min}_k \\
    E^\mathrm{cons}_{k,l} - E_k(t) + E^\mathrm{min}_k, & \text{if } E_k(t) < E^\mathrm{cons}_{k,l} + E^\mathrm{min}_k, \\
    \end{cases}
\end{equation}
which corresponds to the truck agent checking if the SoE before starting a trip is sufficient to cover the energy consumption of its subsequent tour. If it is, the truck does not charge. If, conversely, $E_k(t) < E^\mathrm{cons}_{k,l}$, then the truck charges $E^\mathrm{cons}_{k,l} - E_k(t) + E^\mathrm{min}_k$, where $E^\mathrm{min}_k$ is a safety factor. For the time to start charging $b_k$ we define
\begin{equation}
    \label{eq:abm1}
    b_k =
    \begin{cases}
    \text{No Charge}, & \text{if } E_k(t) \geq E^\mathrm{cons}_{k,l} + E^\mathrm{min}_k \\
    t, & \text{if } E_k(t) < E^\mathrm{cons}_{k,l} + E^\mathrm{min}_k, \\
    \end{cases}
\end{equation}
which corresponds to an immediate start of charge if it is required. For the choice of charger type we have
\begin{equation}
    \label{eq:abm3}
    c_k =
    \begin{cases}
    \text{No Charge}, & \text{if } E_k(t) \geq E^\mathrm{cons}_{k,l} + E^\mathrm{min}_k \\
    \text{ChargerChoice}(R, Q), & \text{if } E_k(t) < E^\mathrm{cons}_{k,l} + E^\mathrm{min}_k, \\
    \end{cases}
\end{equation}
where trucks are assumed to choose the highest power rating charger that has the smallest queue following the $\text{ChargerChoice}(R, Q)$ Algorithm (see Algorithm~\ref{algo:abm:choice}). For the power to be drawn during the charging session we define
\begin{equation}
    \label{eq:abm4}
    d_k =
    \begin{cases}
    \text{0}, & \text{if } E_k(t) \geq E^\mathrm{cons}_{k,l} + E^\mathrm{min}_k \\
    e^\mathrm{char}_{r=c_k}, & \text{if } E_k(t) < E^\mathrm{cons}_{k,l} + E^\mathrm{min}_k, \\
    \end{cases}
\end{equation}
which refers to vehicles charging at the maximum available power provided by the selected charger.
We denote the rule-based decision making with the policies described in eqs.~\eqref{eq:abm1}-\eqref{eq:abm3} as $\pi^{\mathrm{rule}}_k=[a^{\mathrm{rule}}_k,b^{\mathrm{rule}}_k,c^{\mathrm{rule}}_k,d^{\mathrm{rule}}_k]$.
Alternatively, trucks can follow a given charging schedule resulting from the solution of Problem 1, which we denote $\pi^{\mathrm{opt}}_k=[a_k^\mathrm{opt},b_k^\mathrm{opt},c_k^\mathrm{opt},d_k^\mathrm{opt}]$.

\begin{algorithm}
    \caption{Charger Choice Logic}
    \label{algo:abm:choice}
    \begin{algorithmic}[1]
        \scriptsize
    \Require Set of charger types $R$, queues $Q_r$
    \Ensure Selected charger $c_k$
    \Procedure{ChargerChoice}{$R$, $Q_r$}
        \Comment{Selects the fastest from possible charger types $R$ based on their queues $Q$}
        \State $cand \gets Q.\text{get}(Q.\text{size()} - 1)$
        \State $c_k \gets \text{null}$ \Comment{Initialize selected charger}

        \For{$j \gets R.\text{size()} - 1$; $j > -1$; $j \gets j - 1$}
            \If{$cand > Q.\text{get}(j)$}
                \State $cand \gets Q.\text{get}(j)$
                \State $c_k \gets R.\text{get}(j)$
            \EndIf
        \EndFor

        \State \textbf{return} $c_k$
    \EndProcedure
    \end{algorithmic}
\end{algorithm}
\subsubsection{Distribution Centers and Retailers}
\begin{figure}[!t]
\begin{centering}
\includegraphics[width=\linewidth]{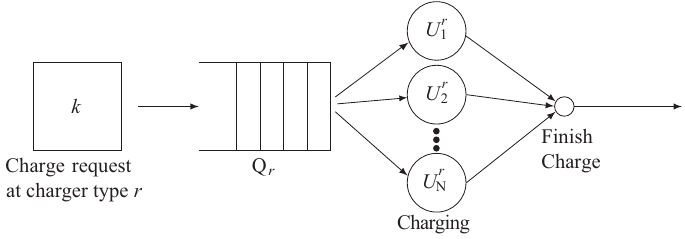}
\caption{
Charging queuing process. Upon entrance each truck agent $k$ enters a FIFO queue $Q_r$ associated with each charger type $r$. This can be pre-determined from the optimization module ($c_k^\mathrm{opt}$) or decided on arrival through the vehicle policy ($c_k^\mathrm{rule}$). Once a charger $U^r_n$ becomes available, the truck is held until the energy required is complete, at which point $k$ is released from the process.
}
\label{fig:queuing-abm}
\end{centering}
\end{figure}

The DCs and retailers indexed by $i$ are located at their respective latitude and longitude, and can therefore be queried in the simulation model to obtain travel times and shortest path routing between each pair.
In this paper, the loading and unloading duration and transportation times between locations are all given in the initial data from Section~\ref{sec:meth:intro}.
The main role of the locations is the provision of charging infrastructure for the vehicles, which enter an embedded queuing model (where each charger type $r$ has its respective queue $Q_r$) and charge when an outlet becomes available, depicted in Fig.\ref{fig:queuing-abm}.
Locations have the agency to stop a vehicle from accessing chargers if a limit connection power $P^\mathrm{max}_i$ is reached, this is enforced with the control policy $\pi_i$ executed each time a vehicle requests to exit the queue $Q_r$ and access a charger $U_n^r$, and every time a vehicle completes its charge.
\( \pi_i \) checks if the total power in use \( P_i(t) \), combined with the power demand of the arriving vehicle \( d_k \), does not exceed the location's maximum contracted power \( P_i^\mathrm{max} \):
\begin{equation}
\label{eq:abmArr}
\pi_i=
\begin{cases}
    \text{ allow to exit } Q_r, & \text{if }  P_i(t) + d_k \leq P_i^\mathrm{max} \\
    \text{ retain in } Q_r, & \text{otherwise},\\
\end{cases}
\end{equation}
thus total power usage remains within the peak contracted power, in line with expected behaviour in privately owned charging stations.
\subsubsection{Output Metrics}
To analyse the performance of our designs, we run $N$ simulations from which we obtain a distribution for relevant metrics.
We have probability distributions for three parameters, the transport duration  $t^\mathrm{tpt}$ , the energy consumption $E^\mathrm{cons}_{k,l}$, and the loading and unloading times $t^\mathrm{load/unload}$.
For the purpose of simplicity, at this stage these are modeled as normally distributed random variables \( X \sim N(x, \delta_x \cdot x) \), where $\delta_x$ is the coefficient of variation of each parameter ($\delta_x = \frac{\sigma_x}{\mu_x}$). Averaging results across simulations, we obtain the following metrics:

\noindent
\textit{Costs}: The total cost of the charging infrastructure required to meet the charging demands of the fleet is obtained as the multiplication of the charger cost and the number of chargers (as given in eq.~\eqref{eq:cost:infra} ). The total energy costs for the operation are obtained by the multiplication of the price of energy and the energy charged at each simulation time step, and then averaged  $\overline{C^\mathrm{energy}} =  \frac{1}{N}\cdot \sum_{k \in \mathcal{K}, l \in \mathcal{L}_k}  c^\mathrm{energy}_{k,l} $.

\noindent
\textit{Reliability}: We track two main indicators of reliability: First, the probability of a truck SoE ($E_k$) falling below the threshold $E^\mathrm{min}_k$ in a given trip, approximated by the average number of times the trucks reach the condition $E(t)<E^\mathrm{min}_k$ during the simulation, averaged over all trucks: $P(E_k < E_k^\mathrm{min}) \approx \frac{1}{N} \cdot \sum_{k \in \mathcal{K}, l \in \mathcal{L}_k}  \mathds{1}_{\{E_k\leq E_k^\mathrm{min}\}}$.
Second, we record the average deviation from scheduled arrival times ($t^\mathrm{arr.act}_{k,l}$), obtained by averaging the deviations for all trucks and all deliveries carried out: $\overline{t^\mathrm{delay}} =  \frac{1}{N \cdot K}\cdot \sum_{k \in \mathcal{K}, l \in \mathcal{L}_k}  (t^\mathrm{arr}_{k,l}-t^\mathrm{arr.act}_{k,l}) $

\noindent
\textit{Waiting Times}: we calculate the average time spent queuing for an available charger and the time spent charging at a CS by averaging the time spent at each respective state over all vehicles and charging instances: $\overline{t^\mathrm{queue}} = \frac{1}{N \cdot K}\cdot\sum_{k \in \mathcal{K}, l \in \mathcal{L}_k}  t^\mathrm{queue}_{k,l} $

\noindent
\textit{Grid requirements}: We compare the instantaneous power ($P_i(t)$) drawn at location $i$ at time $t$ with the contracted power $P^{\mathrm{max}}_i$. We then determine how much time is spent within a specified threshold $\Delta\cdot P^{\mathrm{max}}_i$ for each simulation, and average across $N$:
$U_i(\mathrm{\Delta},P^{\mathrm{max}}_i)=\frac{1}{N} \sum_{t \in \mathcal{T}} \mathds{1}_{P(t) \geq \Delta\cdot P^{\mathrm{max}}_i}$

\subsection{Discussion}\label{sec:discussion}
A few comments are in order regarding the optimization and simulation procedures.
\textit{Optimization:}
The itineraries are provided by the operator in accordance to their planned fleet and existing itineraries, which are based on store requirements and transportation management system.
A different assignment, routing, and sizing of the fleet would vary the results significantly, but this expanded problem is outside the scope of this methodology.
Similarly, our approach excludes from the system scope the energy management system, which at this stage is only reflected in the energy prices used.
This methodology design criteria is based on having a modular, practical, approach, where existing logistics are respected and separatedly analyzed from energy considerations.
Assignments not performable by current commercial vehicles are removed before running the optimization, these could be however, detected and filtered through the use of a binary variable and adding this result in our optimization framework. Similarly, vehicle battery sizes can be added in a straightforward manner.
The optimization problem formulated is a MILP and has worst-case exponential complexity. While this is manageable for our real-world weekly planning scenario involving a large BET fleet, scenarios with significantly larger time-frames or fleets may require more tailored (potentially heuristic) approaches.
\textit{Simulation:} We implement specific policies for both the charging station and the trucks.
Each vehicle is programmed to choose the fastest available charger based on real-time queue data upon arrival. While this may not be the best choice for battery durability, it reflects actual driver behavior where delays are minimized by charging as quickly as possible and avoiding queues.
On the charging station side, the policy is limited to preventing the consumption of power beyond contracted limits.
Although more sophisticated policies could be implemented, for the purpose of this paper, we aim to imitate the current behavior at DCs.
For our stochastic analysis, we focus on variables that are uncertain yet relevant to our stakeholders, keeping the selection limited to ensure straightforward comparisons across experiments.

\section{Case Study}\label{sec:case}
We analyze a real scenario for a main DC in the Northern Netherlands.
We consider a realistic operation with a fleet of 100 vehicles, with three different types of trucks that fulfill the given itineraries: 20 rigid trucks, 50 euro trailers, and 30 city trailers. The main difference among these are their empty weight (10, 14, \unit[14.5]{ton}), maximum loads (8, 10, \unit[10]{ton}) and battery sizes (225, 315, \unit[315]{kWh}, respectively). The starting and ending states of charge of every vehicle are set at 80\% of $E_k^\mathrm{max}$ (i.e., $\phi=0.8$).
The simulated scenario involves 356 locations, with one main DC where almost all departures occur after \unit[3]{AM}. Each location has an average daily tonnage delivery of \unit[29.28]{ton}, and receives an average of 2.15 deliveries per day.
For the charging infrastructure we have 5 different chargers with charging powers of 60, 180, 360, 720 and \unit[1080]{kW}, in accordance to project partner specifications. These have (scaled) initial costs $C^r$ of 0.08, 0.16, 0.33, 0.5 and 1, respectively, and efficiencies $\eta^r$ of 0.98, 0.98, 0.97, 0.97 and 0.97\footnote{Actual initial costs are not given due to confidentiality agreements with project partners}.
Energy prices were obtained for the simulated week (6/11/2023 - 13/11/2023) and vary between 0.043€ and 0.151€ per \unit{kWh}\footnote{Source: https://transparency.entsoe.eu/}.
For easier analysis, and in agreement with stakeholders' current requirements, the charging infrastructure is limited to be located at the DC. Only itineraries that can be feasibly executed using rule-based designs and existing BET technology are considered.
The rule-based design stems from a truck-to-charger ratio of 5 vehicles per charger, and the power ratings are based on planned investments.
To parse and solve Problem 1, we use Yalmip \cite{Loefberg2004} and Gurobi 10.1 on a system with 16 GB of RAM and an Intel i7-9750H processor (20 cores). Optimization solutions were obtained within a global optimality gap of less than 1\% within 4 hours of running time.
For the simulation a test 7-day itinerary is used as outlined in Fig.~\ref*{fig:methodology-process-flow}.
One thousand simulations are performed ($N=1000$) with each simulation run taking in the order of 2 seconds to complete.
To test the robustness of the solutions to different conditions, we let $\alpha^{\mathrm{peak}}$ vary within the values 1 and 2, respectively, to simulate a future current harsher penalization on peak usage. We let $\beta^{\mathrm{peak}}$ vary between 15 and 30 minutes, respectively, to represent the operator's flexibility in schedules within feasible ranges for the operations. We let $\gamma^{\mathrm{energy}}$ vary between 1 and 2, respectively, to represent possible significant increases in energy prices.

\section{Results}\label{sec:res}
The following section displays the initial results obtained from the optimization problem in terms of final design (Table~\ref{tab:infraDesignResults}), costs (Fig.~\ref{fig:res:costs}), and power curves at the DC (Fig.~\ref{fig:res:grid-reqs}). Then the operational results (Table~\ref{tab:summary}) and final costs (Table~\ref{tab:res:energy-costs-sim}) obtained in the agent-based simulation are provided.

\subsection{Optimization}
As shown in Fig.~\ref{tab:infraDesignResults}, the optimized design uses fewer chargers overall (between 7 and 8 for $\alpha^\mathrm{peak}=2$, and 10 for $\alpha^\mathrm{peak}=1$) compared to the rule-based design, which requires 25 chargers in all cases.
The total combined power is also lower (between 5.2 and \unit[3.8]{MW} for $\alpha^\mathrm{peak}=1$, and 3.4 to \unit[3.2]{MW} for $\alpha^\mathrm{peak}=2$) relative to the rule-based design (\unit[4.9]{MW}) .
\begin{table}[!ht]
    \caption{Infrastructure results for varying values of peak factor ($\alpha^\mathrm{peak}$), time slack ($\beta^\mathrm{slack}$), and price multiplier ($\gamma^\mathrm{energy}$) values.}
    \label{tab:infraDesignResults}
    \centering
    \resizebox{\linewidth}{!}{%
    \begin{tabular}{llllccccc}
    \hline
    \multicolumn{1}{c}{\textbf{Design }} & \multicolumn{3}{c}{\textbf{Optimization Parameters}} & \multicolumn{5}{c}{\textbf{Charger Power (kW)}} \\
    \multicolumn{1}{p{1.5cm}}{\raggedright \textbf{ }} &
    \multicolumn{1}{p{0.6cm}}{\raggedright \textbf{$\gamma^\mathrm{energy}$}} &
    \multicolumn{1}{p{0.6cm}}{\raggedright \textbf{$\alpha^\mathrm{peak}$}} &
    \multicolumn{1}{p{0.6cm}}{\raggedright \textbf{$\beta^\mathrm{slack}$}} &
    \multicolumn{1}{p{0.5cm}}{\centering \textbf{60}} &
    \multicolumn{1}{p{0.5cm}}{\centering \textbf{180}} &
    \multicolumn{1}{p{0.5cm}}{\centering \textbf{360}} &
    \multicolumn{1}{p{0.5cm}}{\centering \textbf{720}} &
    \multicolumn{1}{p{0.8cm}}{\centering \textbf{1{,}080}} \\[3pt]
    \hline
    Rule-based  & All & All & All& 10& 10& 4 & 0 & 1 \\
    Co-design       & 1 & 1 & 15 & 0 & 1 & 5 & 3 & 1 \\
    Co-design       & 1 & 2 & 30 & 0 & 0 & 5 & 2 & 0 \\
    Co-design       & 1 & 1 & 30 & 1 & 1 & 6 & 2 & 0 \\
    Co-design       & 1 & 2 & 15 & 0 & 1 & 5 & 2 & 0 \\ \hline
    Co-design       & 2 & 1 & 15 & 0 & 1 & 6 & 3 & 1 \\
    Co-design       & 2 & 2 & 30 & 0 & 0 & 6 & 2 & 0 \\
    Co-design       & 2 & 1 & 30 & 1 & 3 & 6 & 2 & 0 \\
    Co-design       & 2 & 2 & 15 & 0 & 1 & 5 & 3 & 0 \\ \hline
    \end{tabular}%
    }
    \raggedright
    \textbf{Notes:} Higher values of $\alpha^\mathrm{peak}$ indicate higher cost penalizations on the peak component of the objective. Higher $\beta^\mathrm{slack}$ values indicate higher time allowances for the schedules. Higher $\gamma^\mathrm{energy}$ indicate higher energy prices. In the case of rule-based infrastructure the number of chargers is fixed under all parameter combinations.\\
\end{table}
In terms of the type of chargers selected, the optimal solutions rely more on type 3 and 4 chargers (360 and \unit[720]{kW}, respectively). These decisions are influenced by the peak factor and time slacks in each scenario:
higher peak factors reduce the total power used in the solutions, (for $\beta^\mathrm{slack} = 15$, a \unit[1]{MW} and one \unit[720]{kW} less are used when increasing the peak factor, for $\beta^\mathrm{slack} = 30$ a \unit[60]{kW}, a \unit[180]{kW} and a \unit[360]{kW} less are used).
Similarly, increased time slacks allow for more leeway in charging times, and produce minor reductions in the number and power of required chargers (For $\alpha^\mathrm{peak}=1$, a higher time slack avoids using a \unit[1]{MW} charger and a \unit[720]{kW} charger in lieu of adding a \unit[360]{kW} charger; whereas for $\alpha^\mathrm{peak}=2$ a higher time slack uses one less \unit[180]{kW} charger).
Higher $\gamma^\mathrm{energy}$ values also increase the total installed power. In the case of $\alpha^\mathrm{peak}=1$ the increase of the energy costs adds either a \unit[180]{kW} or a \unit[360]{MW} charger, depending on the slack time allowed. In the case of $\alpha^\mathrm{peak}=2$ the charger types increased are either \unit[360]{kW} or \unit[720]{MW} chargers.
\begin{figure}[!t]
    \begin{centering}
    \includegraphics[width=\linewidth]{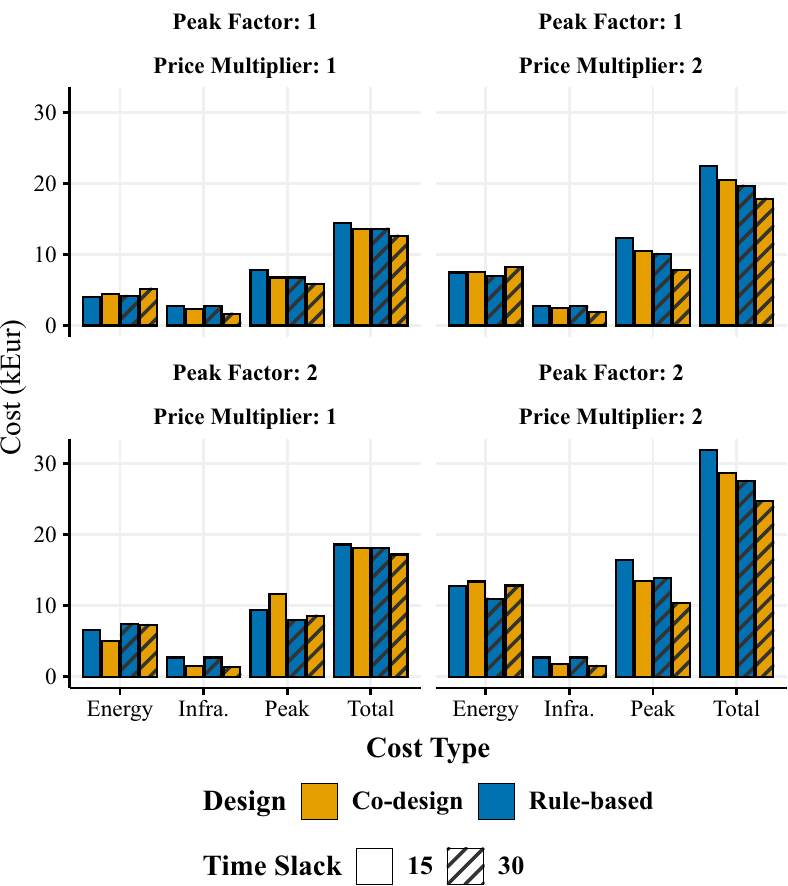}
    \caption{Energy, Infrastructure, Peak and Total costs for each of the experiments. Energy costs correspond to the total energy consumed in charging operations weighted by the energy cost at each time $t$. Infrastructure costs are the combined cost of the charger solution amortized over a 10-year-period. The peak costs are proportional to the absolute maximum power for the period of analysis. Total costs are the sum of infrastructure, energy and peak costs.}
\label{fig:res:costs}
\end{centering}
\end{figure}

Following Fig.~\ref{fig:res:costs}, total costs are lower in the co-designed solutions: The reductions are 6.4\% and 5.2\% for the lower and higher peak factor values, respectively (averaging over $\beta^\mathrm{slack}$).
In general, co-design always yields lower infrastructure costs for this case study (reductions of 48.1\% and 25.9\% ). However, the results for energy and peak costs are mixed. In the case of a lower peak factor ($\alpha^\mathrm{peak}=1$), co-design yields savings in peak costs (reduction of 13.7\%) to the detriment of higher energy costs (increase of 18.2\%).
For the case of a higher peak factor ($\alpha^\mathrm{peak}=2$), co-design yields higher peak costs than the rule-based solution (increase of 23.4\%) but with lower energy costs (decrease of 12.8\%).
The availability of more slack time slack ($\beta^\mathrm{slack}=30$) results in peak, infrastructure and total costs savings (decrease of 18\%, 10.6\%, 5.1\% respectively), but in increases in energy costs (20.6\%), for all design solutions and all peak factors.
Thus the role of higher time slacks is to avoid peaks rather than reducing charging costs.
In general, co-design for this case study focuses in the reduction of infrastructure and peak power costs in lieu of energy costs; this trade-off follows from the common impact on costs that a reduced infrastructure has on maximum peak costs but, in turn, results in reduced flexibility to charge during lower energy cost periods.
\begin{figure}[!t]
\begin{centering}
\includegraphics[width=\linewidth]{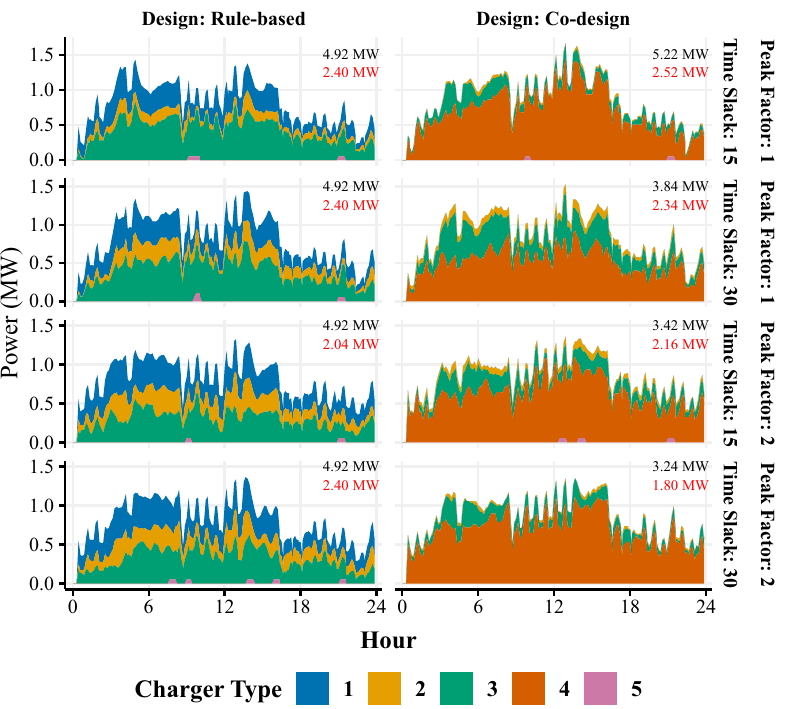}
\caption{Average daily power consumption curves for the DC by charger type. Results are averaged over the simulation time and given for each peak factor, time slack, and design. The red annotation indicates the maximum peak power drawn for the corresponding solution. The black annotation indicates the maximum installed power.}
\label{fig:res:grid-reqs}
\end{centering}
\end{figure}
Fig.~\ref{fig:res:grid-reqs} shows that these different designs have an impact on the resulting charging strategies over the period of analysis. Generally, we can observe that all solutions operate well below the maximum installed power, and the average consumption curve is roughly half of the peak consumption. Co-designed solutions in general have an installed capacity that matches more closely the maximum power drawn.
In all cases the charging is focused on times of lower energy prices constrained to vehicle availability on site. This results in two clear peaks, one around the middle of the day (12:00-17:00) and one in the early morning (03:00-09:00). In the case of co-designed solutions, more power is drawn in the period 09:00 to 12:00.
In the case of rule-based infrastructure, a predominant use of type 3 chargers is noticeable for lower peak factors, with a more even distribution of charger type usage for higher peak factors. The type 5 (\unit[1]{MW}) charger is barely used, only for specific low energy price instances.
For co-designed infrastructure solutions, type 4 (\unit[720]{kW}) chargers are dominating almost all the power transfer for the simulation period. Type 3 (\unit[360]{kW}) chargers are increasingly used with higher time slacks.

\subsection{Agent-based Simulation}

\newcolumntype{L}[1]{>{\raggedright\arraybackslash}p{#1}}
\newcolumntype{C}[1]{>{\centering\arraybackslash}p{#1}}

\begin{table}[ht]
  \caption{Summary statistics for each infrastructure-control combination}
  \label{tab:summary}
  \centering
  \begin{tabularx}{\linewidth}{
      L{0.5cm}%
      C{1.6cm}%
      C{1.55cm}
      C{1.55cm}
      C{1.55cm}
  }
    \hline
    \textbf{Exp.}
      & $\bm{t^{\mathrm{delay}}}$ (min)
      & $\bm{t^{\mathrm{char}}}$ (min)
      & $\bm{t^{\mathrm{que}}}$ (min)
      & $\bm{P_{\mathrm{f}}}(10^{-2})$ \\
    \hline
    \rowcolor{expRR!50}
    RR & -11.3 ± 2.7 & 19.4 ± 4.8  & 4.5 ± 1.0   & 0 ± 0  \\
    \rowcolor{expRO!50}
    RO & -9.8 ± 3.4  & 62.9 ± 14.5 & 0.03 ± 0.04 & 0.3 ± 0.3 \\
    \rowcolor{expOR!50}
    OR & -11.8 ± 2.9 & 14.6 ± 2.8  & 9.2 ± 2.4   & 0 ± 0  \\
    \rowcolor{expOO!50}
    OO & -10.5 ± 3.4 & 16.3 ± 2.6  & 0.3 ± 0.3   & 0.2 ± 0.3\\
    \hline
    \multicolumn{5}{l}{\textit{Note:} All values are reported as the mean ± standard deviation. $N$= 1000.}
  \end{tabularx}
\end{table}

The parameters chosen for validation correspond to the case $\alpha^\mathrm{peak}=2$,  $\beta^\mathrm{slack}=15$ and $\gamma^\mathrm{energy}=1$.
The policy for the vehicles $(\pi_k)$ under rule-based scheduling uses $E_k^\mathrm{min} = 0.05 \cdot E_k^\mathrm{max}$.
The stochastic variables $t^\mathrm{tpt}$, $t^\mathrm{load/unload}$ and $E_{k,l}^\mathrm{cons}$ have a coefficient of variation $\delta_x = 0.05$.
The policy for the locations $(\pi_i)$ uses as a threshold $P_i^\mathrm{max} = P_i^\mathrm{conn}$, where $P_i^\mathrm{conn}$ corresponds to the peak power obtained from the optimization for $\alpha^\mathrm{peak}=2$ and $\beta^\mathrm{slack}=15$.
Operational results averaged per vehicle are shown in Tab.~\ref{tab:summary}, these are given for each of the four combinations depicted in Fig.~\ref{fig:methodology-process-flow}.
The impact of charger design on charging times is patent: Rule-based infrastructure (RO and RR) results in higher charging times compared to the optimized infrastructure (OO and OR), for both optimized and rule-based operations.
Particularly noticeable is the increase of mean charging times (62.9$\pm$\unit[14.5]{min} and 19.4$\pm$\unit[4.8]{min}) by using optimized schedules with rule-based infrastructure.
The difference is reduced with the optimized infrastructure (16.3$\pm$\unit[2.6]{min} and 14.3$\pm$\unit[2.8]{min}), but still shows decreased charging times with rule-based operations.
This difference is due to the greedy choice of faster chargers in rule-based operations compared to a more even distributed use of chargers in optimized operations (also exhibited in the higher spread in charging times for experiment RO).
The impact of optimized schedules can be observed in the queuing times, which are higher for rule-based operations (mean times of 4.5$\pm$\unit[1]{min} and 9.2$\pm$\unit[2.4]{min}).
If there was no uncertainty in the operations, optimized scheduling would have perfect coordination and no queuing times, but with the uncertainties specified, result in still very small queuing times (0.03$\pm$ \unit[0.04]{min} and 0.3$\pm$ \unit[0.3]{min}).
When analyzing realiability results, the impact of online decision making becomes clear.
In terms of failure probability (as defined in Sec.~\ref{section:abm}) the optimized scheduling shows substantially higher failure rates.
For the rule-based infrastructure, the increase in mean failure rates per vehicle-trip for a week increases from less than $1 \cdot 10^{-4}$ to $2.8 \cdot 10^{-3}$, whereas for the optimized infrastructure, the increase is from less than $1 \cdot 10^{-4}$ to $2.3 \cdot 10^{-3}$.
Similarly, schedule delays are higher with optimized scheduling: For the rule-based infrastructure the increase in mean values is from -11.3$\pm$\unit[2.7]{min} to -9.8$\pm$\unit[3.4]{min}; whereas for the optimized infrastructure the increase is from -11.8$\pm$\unit[2.9]{min} to 10.5$\pm$\unit[3.4]{min}.
Additionally, optimized schedules have a higher delay variability.
\begin{table*}[!hptb]
    \caption{Total charging cost and total power utilization rates for the simulated week}
    \label{tab:res:energy-costs-sim}
    \resizebox{\textwidth}{!}{%
    \begin{tabular}{llllllll}
    \hline
    \multicolumn{1}{p{1cm}}{\raggedright \textbf{Design}} &
    \multicolumn{1}{p{1cm}}{\raggedright \textbf{Scheduling}} &
    \multicolumn{1}{p{1.9cm}}{\raggedright \textbf{Energy Cost} \\ \textbf{Simulation (€)}} &
    \multicolumn{1}{p{2.3cm}}{\raggedright \textbf{Energy Cost} \\ \textbf{Optimization (€)}} &
    \multicolumn{1}{p{1.8cm}}{\raggedright \textbf{Weekly Difference (€)}} &
    \multicolumn{1}{p{1.8cm}}{\raggedright \textbf{Yearly Difference (€)}} &
    \multicolumn{1}{p{1.5cm}}{\raggedright \textbf{$\mathbf{U_i(\mathrm{\Delta}=0.5)}$ (\%)}} &
    \multicolumn{1}{p{1.5cm}}{\raggedright \textbf{$\mathbf{U_i(\mathrm{\Delta}=0.85)}$ (\%)}}\\
    \hline
    \rowcolor{expRR!50}
    Rule-based & Rule-Based & 7,651 & 6,605 & 1,046 & 54,392 & 64.6 & 13.5\\
    \rowcolor{expRO!50}
    Rule-based & Optimized  & 6,853 & 6,605 & 248   & 12,896 & 42.9 & 3.2\\
    \rowcolor{expOR!50}
    Co-design  & Rule-Based & 5,899 & 4,956 & 943   & 49,036 & 63.7 & 24.5\\
    \rowcolor{expOO!50}
    Co-design  & Optimized  & 5,087 & 4,956 & 131   & 6,812  & 48.1 & 13.1\\
     \hline
    \end{tabular}%
    }
    \end{table*}
Although rule-based methods show overall acceptable performance, a quick inspection of Table \ref{tab:res:energy-costs-sim} reveals the drawbacks of this approach.
Rule-based energy costs are significantly higher compared to its optimized counterparts, being 10.3\% and 16\% higher for rule-based and co-designed infrastructure, respectively.
This increase stems from the lack of coordination in individual vehicle decision-making with energy prices.
In addition, the drop in performance is higher when using rule-based scheduling with an optimized infrastructure.
Last, rule-based scheduling operates 21.7\% and 15.6\% more time over a threshold of $0.5 \cdot P^\mathrm{max}$, and more critically, 10.3\% and 11.4\% more at a threshold over $0.85 \cdot P^\mathrm{max}$.
This means the charging infrastructure is used closer to capacity for larger periods of time when using rule-based scheduling, revealing an operation closer to congestion, with possible negative implications in station operations and in case of other contingencies.

\subsection{Discussion}\label{sec:discussion-res}
An immediate result from this work is that depot-based itineraries from major retail operators in the Netherlands can be fulfilled with commercially available BET charging technologies, in agreement with the findings of \cite{Whitehead.etal2022} and \cite{spethDepotSlowCharging2024}.
BETs with battery capacities in the 250-\unit[300]{kWh} range, consumptions per \unit{km} of~\unit[1-1.4]{kWh/km} and chargers in the \unit[60-720]{kW} power range, with local grid limitations below \unit[3]{MW}, are compatible with current logistic itineraries, without support of local microgrids.
The results support the procurement of higher powered chargers, not only for meeting tight schedules, but also for long-term energy cost savings.
For this case, the chargers mostly used by co-designed solutions are 360 and \unit[720]{kW} chargers, while \unit[1]{MW} charging are not a strict requirement in this use-case.
This is in alignment with the results from \cite{spethDepotSlowCharging2024,Mishra.etal2022}, but with the caveat that use of higher powered chargers at depots can lead to lower costs.
Particularly relevant will be the use of charging scheduling systems: Even when performing under uncertain conditions of travel time and energy consumption, these deliver significant benefits in terms of energy and peak costs, result in lower impacts on the grid connection, and make better use of the available infrastructure.
Although rule-based strategies may result in acceptable queuing times, they perform significantly worse in terms of costs.
Nevertheless, active control along with schedulers will be key to mitigate existing operational uncertainties in logistics.

\section{Conclusion}\label{sec:dis}
In this paper, we present a framework designed to jointly optimize the charging infrastructure and schedules for Battery Electric Trucks (BETs).
We validated this framework using real-world data in agent-based simulations that reflect realistic operational decision-making.
We find jointly optimizing charging infrastructure and delivery schedules can significantly reduce total costs (between 5.2\% to 6.4\%) and installed power (20.1\% on average) compared to rule-based baseline.
Without co-design, even an optimized charging schedule can lead to higher long-term operational costs (24.9\% average decrease for the simulation dataset).
The charging infrastructure design is influenced by energy prices, though it exhibits limited sensitivity. However, total cost differences compared to rule-based designs become more pronounced under higher energy price scenarios.
We also find optimizing operations is crucial to navigate peak power limitations (10.3\% to 11.4\% less time with usage levels above 85\% peak power usage).
Furthermore, online control can help mitigate uncertainties in consumption and transport times (average reduction of 100\% in failures and 12.2\% in delay times) as evidenced with rule-based charging scheduling.
We believe that explainable optimization and simulation frameworks, equipped with realistic operation rules and reliability metrics, are essential for improving decision-making and facilitating a smoother transition to BETs.
Future work will expand this framework to include vehicle assignment and routing, online vehicle and charging control, and energy management.
Additionally, we will analyze how different pricing and operational constraints impact final designs in collaboration with project partners.

\addtolength{\textheight}{0cm}%

\section*{Acknowledgment}

We thank Dr.~I.~New, Ir.~M.~Clemente and Dr.~F.~Paparella for proofreading this paper. This publication is part of the project GTD-Elektrifikatie, made possible by the Ministry of Economic Affairs and Climate Policy of the Netherlands.

\enlargethispage{2\baselineskip}


\begin{thebibliography}{10}
\providecommand{\url}[1]{#1}
\csname url@samestyle\endcsname
\providecommand{\newblock}{\relax}
\providecommand{\bibinfo}[2]{#2}
\providecommand{\BIBentrySTDinterwordspacing}{\spaceskip=0pt\relax}
\providecommand{\BIBentryALTinterwordstretchfactor}{4}
\providecommand{\BIBentryALTinterwordspacing}{\spaceskip=\fontdimen2\font plus
\BIBentryALTinterwordstretchfactor\fontdimen3\font minus
  \fontdimen4\font\relax}
\providecommand{\BIBforeignlanguage}[2]{{%
\expandafter\ifx\csname l@#1\endcsname\relax
\typeout{** WARNING: IEEEtran.bst: No hyphenation pattern has been}%
\typeout{** loaded for the language `#1'. Using the pattern for}%
\typeout{** the default language instead.}%
\else
\language=\csname l@#1\endcsname
\fi
#2}}
\providecommand{\BIBdecl}{\relax}
\BIBdecl

\bibitem{teter2017future}
{International Energy Agency}. (2017) The future of trucks: {{Implications}}
  for energy and the environment. {Available at
  }\url{https://doi.org/10.1787/9789264279452-en}.

\bibitem{burgmeijer2018}
{TNO}. (2018) Research program 2019 logistics and mobility. {Available at
  }\url{https://www.tno.nl/publish/pages/4147/thema_htsm_vp-plannen_2019.pdf}.

\bibitem{itf2021}
{International Transport Forum}. (2021) {{ITF}} transport outlook 2021.
  {Available at }\url{https://doi.org/10.1787/16826a30-en}.

\bibitem{cbsopendata}
{Centraal Bureau voor de Statistiek}. (2023) {{CBS Open}} data {{StatLine}}.
  {Available at }\url{https://opendata.cbs.nl/statline/portal.html}.

\bibitem{Rijksoverheid2021Nieuwe}
{Rijksoverheid}. (2021) Nieuwe afspraken om steden te bevoorraden zonder
  {{CO2-uitstoot}}. {Available at
  }\url{https://www.rijksoverheid.nl/actueel/nieuws/2021/02/09/nieuwe-afspraken-om-steden-te-bevoorraden-zonder-co2-uitstoot}.

\bibitem{Lantz.Joelsson2023}
M.~Lantz and Y.~Joelsson. (2023) Electric heavy-duty trucks - {{Policy
  Outlook}}: {{Planned}} and implemented policies to support battery electric
  heavy-duty vehicles in {{Sweden}}, {{Austria}}, {{Germany}}, the
  {{Netherlands}}, {{UK}} and {{California}} ({{US}}). {Available at
  }\url{https://portal.research.lu.se/files/138033720/Report_129.pdf}.

\bibitem{Ploetz2022}
P.~Pl\"otz, ``Hydrogen technology is unlikely to play a major role in
  sustainable road transport,'' \emph{{Nature Electronics}}, vol.~5, pp. 8--10,
  2022.

\bibitem{galloElectricTruckBus2016}
J.~B. Gallo, ``Electric {{Truck}} \& {{Bus Grid Integration}},
  {{Opportunities}}, {{Challenges}} \& {{Recommendations}},'' \emph{{World
  Electric Vehicle Journal}}, vol.~8, no.~1, pp. 45--56, 2016.

\bibitem{Speth.etal2022a}
D.~Speth, P.~Pl{\"o}tz, S.~Funke, and E.~Vallarella, ``Public fast charging
  infrastructure for battery electric {{Trucks}}\textemdash a model-based
  network for {{Germany}},'' \emph{{Environmental Research: Infrastructure and
  Sustainability}}, vol.~2, p. 025004, 2022.

\bibitem{Transport&Environment2023}
{European Federation for Transport and Environment}. (2023) Companies ask
  {{EU}} negotiators to boost vehicle charging infrastructure. {Available at
  }\url{https://www.transportenvironment.org/discover/companies-ask-eu-negotiators-to-boost-vehicle-charging-infrastructure/}.

\bibitem{Zhang.etal2021}
M.~Zhang, X.~Zhu, B.~Mather, P.~Kulkani, and A.~Meintz, ``Location
  {{Selection}} of {{Fast-Charging Station}} for {{Heavy-Duty EVs Using GIS}}
  and {{Grid Analysis}},'' in \emph{{IEEE Power \& Energy Society Innovative
  Smart Grid Technologies Conference}}, 2021, pp. 1--5.

\bibitem{Whitehead.etal2022}
J.~Whitehead, J.~Whitehead, M.~Kane, and Z.~Zheng, ``Exploring public charging
  infrastructure requirements for short-haul electric trucks,'' \emph{{Int.\
  Journal of Sustainable Transportation}}, vol.~16, no.~9, pp. 775--791, 2022.

\bibitem{spethDepotSlowCharging2024}
D.~Speth and P.~Pl{\"o}tz, ``Depot slow charging is sufficient for most
  electric trucks in {{Germany}},'' \emph{Transportation Research Part D:
  Transport and Environment}, vol. 128, 2024.

\bibitem{NREL2023}
{National Renewable Energy Lab (NREL)}. (2020) {{EVI-EnSite}}: {{Electric
  Vehicle Infrastructure}} {{Energy Estimation}} and {{Site Optimization
  Tool}}. {Available at
  }\url{https://www.nrel.gov/transportation/evi-ensite.html}.

\bibitem{Mishra.etal2022}
P.~Mishra, E.~Miller, S.~Santhanagopalan, K.~Bennion, and A.~Meintz, ``A
  {{Framework}} to {{Analyze}} the {{Requirements}} of a {{Multiport
  Megawatt-Level Charging Station}} for {{Heavy-Duty Electric Vehicles}},''
  \emph{{Energies}}, vol.~15, no.~10, pp. 3788--3794, 2022.

\bibitem{pelletierChargeSchedulingElectric2018}
S.~Pelletier, O.~Jabali, and G.~Laporte, ``Charge scheduling for electric
  freight vehicles,'' \emph{{Transportation Research Part B: Methodological}},
  vol. 115, pp. 246--269, 2018.

\bibitem{zahringerTimeVsCapacity2022}
M.~Z{\"a}hringer, S.~Wolff, J.~Schneider, G.~Balke, and M.~Lienkamp, ``Time vs.
  {{Capacity}}\textemdash{{The Potential}} of {{Optimal Charging Stop
  Strategies}} for {{Battery Electric Trucks}},'' \emph{{Energies}}, vol.~15,
  no.~19, p. 7137, Sep. 2022.

\bibitem{zahringerOptimizingJourneyDynamic2024}
M.~Z{\"a}hringer, O.~Teichert, G.~Balke, J.~Schneider, and M.~Lienkamp,
  ``Optimizing the {{Journey}}: {{Dynamic Charging Strategies}} for {{Battery
  Electric Trucks}} in {{Long-Haul Transport}},'' \emph{Energies}, vol.~17,
  2024.

\bibitem{bragin-joint}
M.~A. Bragin, Z.~Ye, and N.~Yu. (2023) Toward {{Efficient Transportation
  Electrification}} of {{Heavy-Duty Trucks}}: {{Joint Scheduling}} of {{Truck
  Routing}} and {{Charging}}. {Available at
  }\url{https://arxiv.org/abs/2302.00240}.

\bibitem{wangOptimalDispatchRouting2023a}
R.~Wang, T.~Zeng, P.~Keyantuo, J.~Sandoval, A.~Vishwanath, H.~Borhan, and
  S.~Moura, ``Optimal {{Dispatch}} and {{Routing}} of {{Electrified Heavy-Duty
  Truck Fleets}}: {{A Case Study}} with {{Fleet Data}},'' in \emph{{American
  Control Conference}}, 2023, pp. 1729--1734.

\bibitem{BertucciHofmanEtAl2024}
J.~Bertucci, T.~Hofman, and M.~Salazar, ``Joint optimization of charging
  infrastructure placement and operational schedules for a fleet of battery
  electric trucks,'' in \emph{{American Control Conference}}, 2024, available
  online at \url{https://arxiv.org/abs/2310.02181}.

\bibitem{BorsboomFahdzyanaEtAl2021}
O.~Borsboom, C.~A. Fahdzyana, T.~Hofman, and M.~Salazar, ``A convex
  optimization framework for minimum lap time design and control of electric
  race cars,'' \emph{{IEEE Transactions on Vehicular Technology}}, vol.~70,
  no.~9, pp. 8478--8489, 2021.

\bibitem{borshchev2014multi}
A.~Borshchev, ``Multi-method modelling: Anylogic,'' \emph{Discrete-event
  simulation and system dynamics for management decision making}, pp. 248--279,
  2014.

\bibitem{Loefberg2004}
J.~L{\"o}fberg, ``{YALMIP} : A toolbox for modeling and optimization in
  {MATLAB},'' in \emph{{IEEE Int.\ Symp.\ on Computer Aided Control Systems
  Design}}, 2004.

\end{thebibliography}
\end{document}